\PassOptionsToPackage{table,xcdraw}{xcolor}

\documentclass[conference,a4paper]{APSIPA2021}
\usepackage{array, amsmath, amssymb, amsfonts, bm}
\usepackage{graphicx}
\usepackage{multirow, booktabs}
\usepackage{threeparttable}
\usepackage[backend=biber,style=ieee,]{biblatex}
\usepackage{xintexpr}
\usepackage{flushend}
\usepackage{lipsum}

\usepackage{geometry}
\IEEEoverridecommandlockouts

\newcommand{\audio}{{(a)}}
\newcommand{\video}{{(v)}}

\begin{document}

\setlength\abovedisplayskip{4.0mm}
\setlength\belowdisplayskip{4.0mm}

\title{
DOA-Aware Audio-Visual Self-Supervised Learning
for Sound Event Localization and Detection
}

\author{
\authorblockN{
Yoto Fujita\authorrefmark{1}\authorrefmark{2}, %
Yoshiaki Bando\authorrefmark{2}, %
Keisuke Imoto\authorrefmark{3}\authorrefmark{2}, %
Masaki Onishi\authorrefmark{2}, %
and Kazuyoshi Yoshii\authorrefmark{1}}

\authorblockA{
\authorrefmark{1}Graduate School of Informatics, Kyoto University, Japan, 
\{fujita.yoto.26m@st.kyoto-u.ac.jp, yoshii@i.kyoto-u.ac.jp\}}

\authorblockA{
\authorrefmark{2}National Institute of Advanced Industrial Science and Technology, Japan, 
\{y.bando,onishi-masaki\}@aist.go.jp}

\authorblockA{
\authorrefmark{3}Faculty of Science and Engineering, Doshisha University, Japan, 
keisuke.imoto@ieee.org}
}

\maketitle

\begin{abstract}
This paper describes 
 sound event localization and detection (SELD)
 for spatial audio recordings captured 
 by first-order ambisonics (FOA) microphones.
In this task, one may
 train a deep neural network (DNN)
 using FOA data annotated 
 with the classes and directions of arrival (DOAs) of sound events.
However, the performance of this approach
 is severely bounded by the amount of annotated data.
To overcome this limitation,
 we propose a novel method of pretraining 
 the feature extraction part of the DNN 
 in a self-supervised manner.
We use spatial audio-visual recordings
 abundantly available as virtual reality contents.
Assuming that sound objects are concurrently observed
 by the FOA microphones and the omni-directional camera,
 we jointly train audio and visual encoders 
 with contrastive learning 
 such that the audio and visual embeddings 
 of the same recording and DOA are made close.
A key feature of our method
 is that the DOA-wise audio embeddings 
 are \textit{jointly} extracted from the raw audio data, 
 while the DOA-wise visual embeddings are \textit{separately} extracted 
 from the local visual crops centered on the corresponding DOA.
This encourages the latent features of the audio encoder
 to represent both the classes and DOAs of sound events.
The experiment using the DCASE2022 Task 3 dataset of 20 hours
 shows non-annotated audio-visual recordings of 100 hours
 reduced the error score of SELD from 36.4 pts to 34.9 pts.
\end{abstract}


\section{Introduction}
\label{sec:introduction}

Sound event localization and detection (SELD) 
 is a task that aims to estimate the activations, classes, 
 and directions of arrival (DOAs) of sound events 
 from multichannel audio recordings~\cite{politis2020overview}.
It is one of the foundations of computational intelligence
 for understanding acoustic environments.
The current standard approach to this task 
 is to train a deep neural network (DNN) in a supervised manner 
 using pairs of audio recordings
 with ground-truth annotations
 \cite{nguyen2020sequence,adavanne2018sound,shimada2021accdoa}.
In general, it is difficult 
 to collect a sufficient amount of annotated audio data 
 covering a wide range of acoustic environments.

A popular solution to this problem
 is data augmentation~\cite{politis2022starss22,Du_NERCSLIP_task3_report}.
Specifically, 
 one can synthesize multichannel audio data  
 by convolving source signals of various classes
 with room impulse responses (RIRs) that simulate 
 various acoustic environments and DOA conditions.
This approach is known 
 to effectively improve the performance of SELD 
 for many common sound events (e.g., music and speech).
However, SELD for some complex sound events (e.g., wildlife sound)
 remains an open problem
 because it is difficult to isolate and capture 
 their individual sound source signals.

\begin{figure}[t]
    \centering
    \includegraphics[width=.95\hsize]{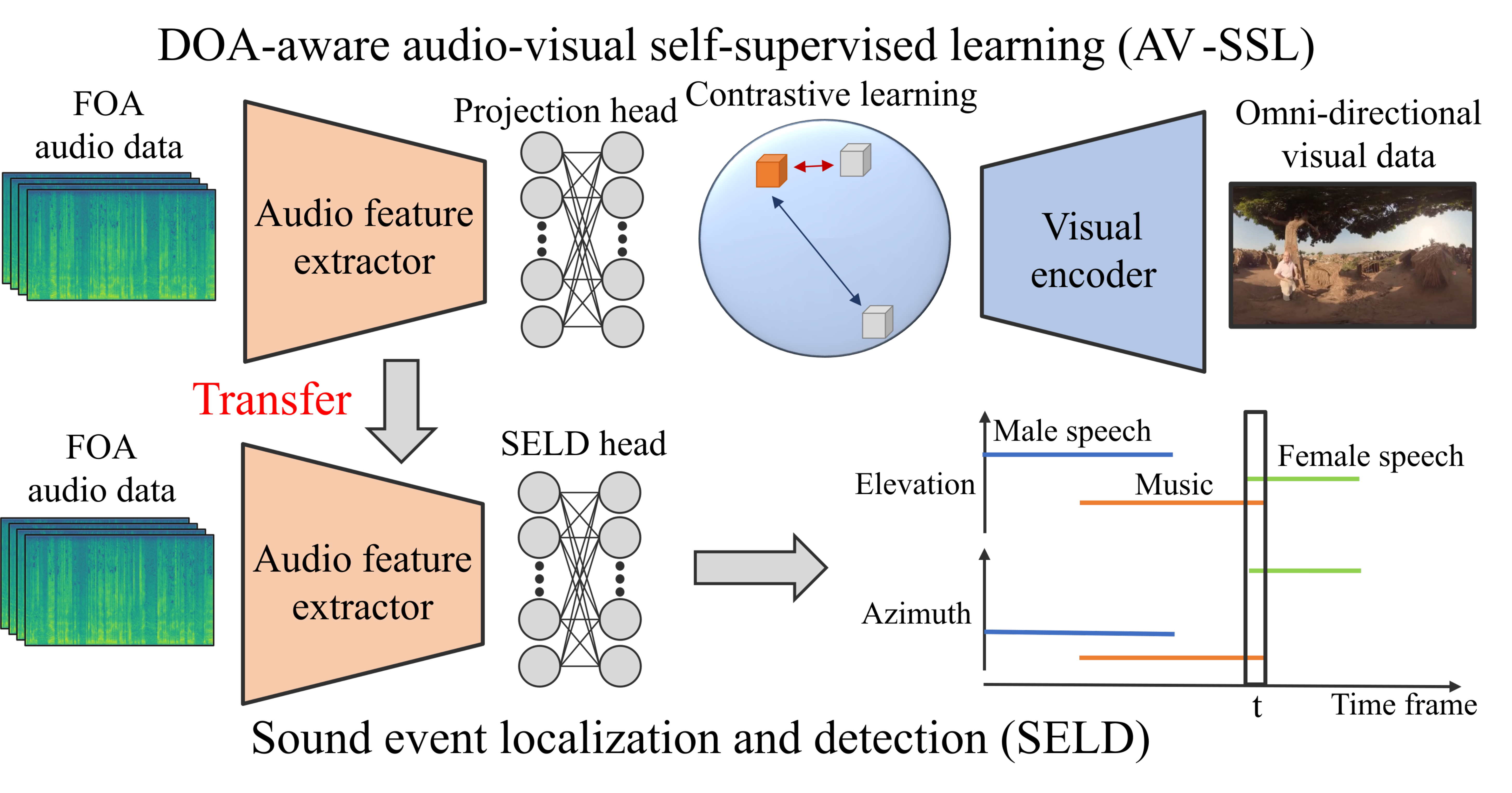}
    \vspace{-1mm}
    \caption{AV-SSL of an audio feature extractor for SELD.}
    \label{fig:AV-SSL}
    \vspace{-4mm}
\end{figure}


Another solution is to pretrain an audio feature extractor 
 that constitutes the front end of a DNN used for SELD 
 with audio-visual self-supervised learning (AV-SSL) 
 (Fig.~\ref{fig:AV-SSL}) \cite{morgado2020learning}.
This approach can make effective use 
 of abundant virtual reality (VR) contents,
 each of which consists of multichannel audio data
 recorded by first-order ambisonics (FOA) microphones
 and 360$^\circ$ equirectangular visual data
 recorded by an omni-directional camera.
Considering the cross-modal co-occurrence 
  between the sounds and appearances of the same objects,
 one can jointly train audio and visual encoders
 in a contrastive fashion
 such that the audio and visual embeddings are made close to each other
 if they correspond to the same DOA and recording (positive sample)
 and far apart otherwise (negative sample).
The front end of the trained audio encoder
 is then used for initializing the audio feature extractor for SELD.

Audio-visual spatial alignment (AVSA)~\cite{morgado2020learning}
 is one of the latest AV-SSL methods.
It takes advantage of the FOA format,
 in which the single-channel audio signal with an arbitrary DOA 
 can be computed from the observed FOA data.
The audio embedding extracted from this enhanced signal 
 are made close to the visual embedding extracted 
 from the visual crop centered on the same DOA 
 in the equirectangular visual data.
However, the audio feature extractor trained in this way
 is insufficient for SELD
 because the DOA features of sound events 
 cannot be extracted from the enhanced signal.
On the other hand, the features useful for DOA estimation of sound events are not learnable with such non-DOA-aware contrastive learning equivalent in principle to AVC~\cite{arandjelovic2017look}.

In this paper, 
 we propose a DOA-aware extension of AVSA.
Our method differs from the conventional AVSA \cite{morgado2020learning}
 in that it jointly extracts 
 audio embeddings over a DOA grid
 from raw FOA audio data
 without DOA-wise signal enhancement.
This encourages the latent features of the audio encoder
 to represent both the classes and DOAs of sound events. 
In addition, this paper also tackles one of the remaining challenges in AVSA
 that the cross-modal co-occurrence 
 between sound and appearance 
 does not hold 
 for visible silent objects and occluded sound objects.
In general, meaningful sound events exist 
 only in a small number of DOAs,
 which limits this \textit{local} contrastive learning
 based on the DOA-wise similarity 
 between audio and visual embeddings.

To mitigate this problem,
 we also propose \textit{global} contrastive learning 
 based on recording-wise audio-visual similarity 
 obtained by averaging the DOA-wise similarities
 over all the DOAs.
Our goal is to maximize the similarity
 when the audio and visual embeddings 
 correspond to the same recording (positive)
 and minimize it otherwise (negative).
To encourage the audio encoder
 to extract DOA information as the latent features,
 we introduce a data augmentation technique
 that randomly spatially rotates only the equirectangular visual data
 to generate negative samples 
 from the same recording.

\begin{figure*}
    \centering
    \includegraphics[width=.85\hsize]{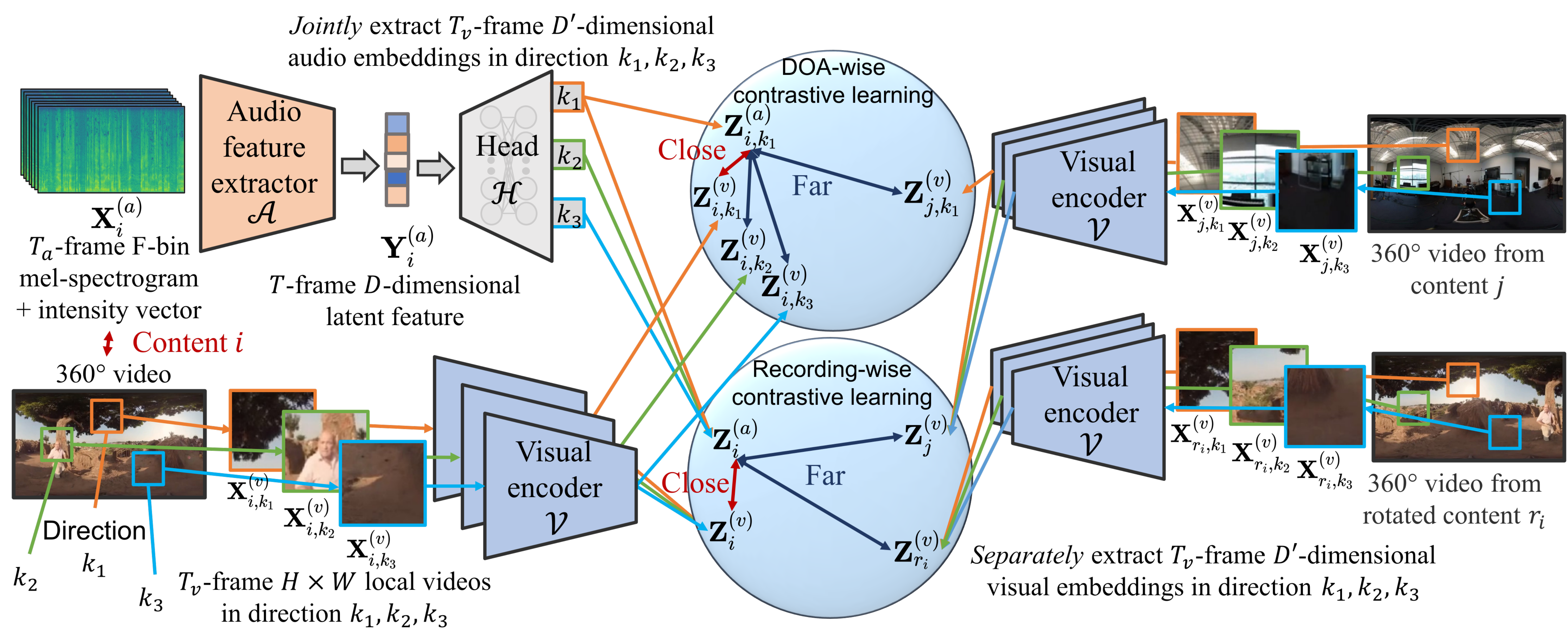}
    \vspace{-2mm}
    \caption{The proposed DOA-wise and recording-wise DOA-aware contrastive learnings for pretraining the audio feature extractor.}
    \label{fig:overview}
    \vspace{-1.5mm}
\end{figure*}

\section{Related Work}

This section reviews 
 existing SELD and AV-SSL methods. 

\subsection{Sound event localization and detection (SELD)}

The modern approach to SELD is 
 to train a DNN in a supervised manner.
For instance,
 the activations of sound events are estimated with a DNN,
 while the DOAs are estimated geometrically \cite{nguyen2020sequence}. 
Both the activations and DOAs 
 can be estimated using a DNN~\cite{adavanne2018sound}. 
Recently, an end-to-end approach to SELD has been proposed
 for directly estimating a DOA vector
 whose length corresponds to the duration of sound events
 in the Cartesian coordinate system~\cite{shimada2021accdoa}.

To improve the robustness of SELD 
 with a limited amount of annotated audio data, 
 data augmentation tequniques
 such as SpecAugment~\cite{park2019specaugment} and Mixup~\cite{takahashi2016deep}
 can be employed.
Particularly for audio data provided in the FOA format,
 rotation-based augmentation
 is known to be effective~\cite{mazzon2019first}.
Multichannel audio data with ground-truth annotations can be synthesized
 by convolving audio samples of various classes with arbitrary transfer functions 
 \cite{politis2022starss22,DBLP:journals/corr/abs-2101-02919}.
Alternatively, a general-purpose audio model~\cite{gong2022ssast} 
 pretrained with a large dataset of audio signals 
 called AudioSet~\cite{gemmeke2017audio} 
 in an unsupervised manner
 can be used for sound event detection~\cite{scheibler2022sound}. 

\subsection{Audio-visual self-supervised learning (AV-SSL)}

At the heart of AV-SSL is contrastive learning, 
 which aims to train audio and visual encoders 
 based on the cross-modal co-occurrence between
 the sounds and appearances of objects.
The information encoded in the audio and visual embeddings may vary 
 depending on the design of positive and negative samples.
AVC~\cite{arandjelovic2017look}, for example,
 uses standard video recordings
 and considers a pair of audio and visual data from the same recording
 as a positive sample and a pair of those from different recordings
 as a negative pair.
The audio and visual embeddings are thus encouraged
 to represent the features of sound event classes. 
AVSA~\cite{morgado2020learning} uses spatial video recordings,
 which are originally made for VR application
 that requires DOA-based audio-visual rendering.
AVSA, however, is essentially identical to AVC
 because it can be regarded as AVC
 for \textit{non-spatial} video standard recordings
 obtained by decomposing \textit{spatial} video recordings
 in a DOA-wise manner.
 
Audio-visual temporal synchronization (AVTS)~\cite{korbar2018cooperative} 
 is another AVC-like self-supervised method.
A key feature of AVTS is that 
 it incorporates audio and visual data in the same recording, 
 but at different moments in time, to negative pairs.
The embeddings are thus encouraged 
 to represent both the classes and activations of sound events. 

\section{Proposed Method}

This section describes two variants of DOA-aware AV-SSL 
 to improve SELD.  
The first variant employs DOA-wise contrastive learning,
 and the other employs recording-wise contrastive learning (Fig.~\ref{fig:overview}).
In both variants, an audio feature extractor $\mathcal{A}$ 
 is used to transform raw FOA data into the latent audio features
 that represent sound event classes and DOAs.
In this study, $K$ discrete points on the Fibonacci lattice~\cite{gonzalez2010measurement} 
 are considered as potential DOAs.
Each DOA $k \in [1, K]$ is defined 
 by an azimuth angle $\theta_k$ 
 and an elevation angle $\phi_k$.
A projection head $\mathcal{H}$ is then used to \textit{jointly} convert
 the latent features
 to the audio embeddings over the DOA grid.
A visual encoder $\mathcal{V}$, in contrast, 
 is used to \textit{separately} convert visual crops 
 of a 360$^\circ$ equirectangular visual data
 over the DOA grid
 to the visual embeddings.
Let $I$ be the total number of recordings. 

In DOA-wise contrastive learning,
 we maximize the local similarity 
 between DOA-wise audio and visual embeddings 
 when they correspond to the same DOA (positive sample),
 and minimize it otherwise (negative sample).
In recording-wise contrastive learning, in contrast,
 we maximize global similarity 
 obtained by averaging the local similarities 
 over the DOA grid
 when the audio and visual embeddings correspond to the same recording (positive),
 and minimize it otherwise (negative).

Once $\mathcal{A}$, $\mathcal{H}$, and $\mathcal{V}$ are trained jointly 
 in a self-supervised manner,
 $\mathcal{A}$ is connected to another head $\mathcal{H}'$ for SELD
 and the entire network is fine-tuned in a supervised manner.

\subsection{Audio encoding}\label{AA}

Let $\mathbf{X}_{i}^\audio \in \mathbb{R}^{(4+3) \times F \times T_a}$
 be the multichannel audio spectrogram of spatial recording $i \in [1, I]$
 obtained by stacking the mel spectrograms 
 of the four-channel audio signals of the FOA format
 and the three intensity spectrograms on the orthogonal axes, 
 where $F$ is the number of mel frequency bins,
 and $T_a$ is the number of frames.
A series of frame-wise audio latent features denoted by
 $\mathbf{Y}_i^\audio \triangleq \{ \mathbf{y}^\audio_{it} \}_{t = 1}^{T_a}$
 is obtained with the audio feature extractor $\mathcal{A}$ as follows:
\begin{align}
    \mathbf{Y}_i^\audio 
    \leftarrow 
    \mathcal{A}\left(\mathbf{X}_{i}^\audio\right).
\end{align}
A series of DOA- and frame-wise audio embeddings 
 $\mathbf{Z}^\audio_{i} \triangleq \{ \mathbf{z}^\audio_{ikt} \}_{k=1,t=1}^{K,T_v}$
 is then obtained 
 with the projection head $\mathcal{H}$:
\begin{align}
    \{\mathbf{z}^\audio_{ikt}\}_{k=1}^K 
    \leftarrow 
    \mathcal{H}\left(\mathbf{y}^\audio_{it}\right),
\end{align}
where the features $\mathbf{y}^\audio_{it}$ at each frame $t$
 are independently transformed for temporally localizing information,
 followed by an adaptive average pooling
 to match the visual data of $T_v$ frames.

\subsection{Visual encoding}

Let $\mathbf{X}_{i}^\video 
 \triangleq \{ \mathbf{x}^\video_{ikt} \}_{k=1,t=1}^{K,T_v}$
 be the series of DOA-wise local images
 cropped from the 360$^\circ$ equirectangular visual data
 of recording $i \in [1, I]$,
 where $\mathbf{x}^\video_{ikt} \in \mathbb{R}^{H \times W}$ 
 is a local image that corresponds to DOA $k \in [1, K]$ 
 at time $t \in [1, T]$,
 $H$ and $W$ are the height and width of each image
 (cropping size),
 and $T_v$ is the number of frames.
Note that 
 $\mathbf{x}^\video_{ikt}$ is centered on DOA $k$
 on the Gnomonic projection~\cite{weisstein2001gnomonic} 
 of the original spatial visual data.
A series of DOA- and frame-wise visual embeddings
 $\mathbf{Z}_i^\video \triangleq \{ \mathbf{z}^\video_{ikt} \}_{k=1,t=1}^{K,T_v}$
 is obtained with the visual encoder $\mathcal{V}$ as follows:
\begin{align}
    \{\mathbf{z}^\video_{ikt}\}_{t=1}^{T_v}
    \leftarrow
    \mathcal{V}\left(\{\mathbf{x}^\video_{ikt}\}_{t=1}^{T_v}\right),
\end{align}
where the same visual encoding is independently applied 
 to each DOA $k$ to obtain the embeddings 
 that represent the classes of visible sound objects.

\subsection{Self-supervised learning (pretraining)}

We describe the two variants of contrastive learning.

\subsubsection{Similarity measures}

Since the audio and visual data of the same spatial recording
 usually have the DOA-wise correspondence,
 we define the similarity
 between recordings $i$ and $j$ for each DOA $k$ 
 as the cosine similarity as follows:
\begin{align}
    \mathcal{S}_\mathrm{DOA}(\mathbf{Z}^\audio_{ik}, \mathbf{Z}^\video_{jk}) 
    &= 
    \frac{1}{T_v}\sum_{t=1}^{T_v} 
    \frac{
    \mathbf{z}^{\audio \mathsf{T}}_{ikt}
    \mathbf{z}^\video_{jkt}
    }
    {
    \|\mathbf{z}^\audio_{ikt}\| 
    \|\mathbf{z}^\video_{jkt}\|
    },
\end{align}
where $\mathbf{Z}^\audio_{ik} \triangleq \{\mathbf{z}^\audio_{ikt}\}_{t=1}^{T_v}$
and $\mathbf{Z}^\video_{ik} \triangleq \{\mathbf{z}^\video_{ikt}\}_{t=1}^{T_v}$.

We here focus on the InfoNCE loss~\cite{oord2018representation} defined as follows:
\begin{align}
    \hspace{-2.8mm}\mathcal{I} \left(
    \mathbf{Z}, \mathbf{Z}_p, \mathbf{U}_n \right) = - \frac{\exp\left(\mathcal{S}(\mathbf{Z}, \mathbf{Z}_p)/\tau\right)}{\sum_{\mathbf{Z}_n \in \mathbf{U}_n} \exp\left(\mathcal{S}(\mathbf{Z}, \mathbf{Z}_n)/\tau\right)},
\end{align}
where $\mathbf{Z}$ is an anchor,
 $\mathbf{Z}_p$ is a positive sample,
 $\mathbf{U}_n$ is a set of negative samples,
 and $\tau$ is a temperature hyperparameter.
Minimizing $\mathcal{I}$ 
 encourages the similarity of $\mathbf{Z}$ to $\mathbf{Z}_p$
 to be larger than to $\mathbf{U}_n$
 in a contrastive manner.

\subsubsection{DOA-wise contrastive learning}

The loss $\mathcal{L}_\mathrm{DOA}$ used in this variant
 is calculated as follows:
\begin{align}
    \hspace{-2.8mm}\mathcal{L}_\mathrm{DOA} 
    = \sum_{i,k=1}^{I,K} \mathcal{I} \left(
    \mathbf{Z}^\audio_{ik}, \mathbf{Z}^\video_{ik}, \{ \mathbf{Z}^\video_{jk} \}_{j=1}^I \cup \{ \mathbf{Z}^\video_{ik'} \}_{k'=1}^K \right),
\end{align}
where the similarity between the audio and visual embeddings 
 is maximized when they correspond to the same DOA.

\subsubsection{Recording-wise contrastive learning}


The loss $\mathcal{L}_\mathrm{REC}$ used in this variant
 is calculated as follows:
\begin{align}
    \mathcal{L}_{\mathrm{REC}} = \sum_{i=1}^I\mathcal{I} \left(
    \mathbf{Z}^\audio_i, \mathbf{Z}^\video_i, \{ \mathbf{Z}^\video_j \}_{j=1}^I \cup \{\mathbf{Z}^\video_{r_i}\} \right),
\end{align}
where $\mathbf{Z}^\video_{r_i}$ is a series of visual embeddings 
 that corresponds to the visual data from the rotated recording $r_i$,
 which is obtained by rotating the spatial information of the recording $i$.
The recording-wise similarity used in $\mathcal{L}_{\mathrm{REC}}$ 
 is obtained by averaging the DOA-wise similarities 
 over all the directions:
\begin{align}
     \mathcal{S}_\mathrm{REC}(\mathbf{Z}^\audio_i, \mathbf{Z}^\video_j) &= \frac{1}{K} \sum_{k = 1}^K \mathcal{S}_\mathrm{DOA}(\mathbf{Z}^\audio_{ik}, \mathbf{Z}^\video_{jk}).
\end{align}

\subsection{Supervised learning (fine-tuning)}

Using annotated data,
 the audio feature extractor $\mathcal{A}$ is fine-tuned for SELD
 based on activity-coupled Cartesian DOA representation (ACCDOA) \cite{shimada2022multi}.
Specifically, multiple sound events of the same class 
 can be separately assigned to different tracks.
Let $\hat{\mathbf{P}} \in \mathbb{R}^{T' \times C \times N \times 3}$ 
 be a multi-ACCDOA vector of $T'$ frames, $C$ classes, $N$ tracks given by
\begin{align}
    \hat{\mathbf{P}} \leftarrow \mathcal{H}'\left(\mathcal{A}\left(\mathbf{X}\right)\right),
\end{align}
where $\mathbf{X} \in \mathbb{R}^{(4+3) \times F \times T'_a}$
 is the multichannel spectrogram computed 
 in the same way as $\mathbf{X}_{i}^\audio$, 
 where $T_a$ is the number of frames.
The projection head $\mathcal{H}'$ consists of 
 several fully-connected layers, 
 followed by an adaptive average pooling 
 to suit the target time resolution $T'$.
Since the SELD label for frame $t$, class $c$, and track $n$ 
 is given as a pair of the activity $a_{tcn}^* \in \{0,1\}$ 
 and the Cartesian DOA vector $\mathbf{R}_{tcn}$, 
 the target ACCDOA vector is given by
\begin{align}
    \mathbf{P}_{tcn}^* = a_{tcn}^* \mathbf{R}_{tcn}.
\end{align}
The overall network is trained to minimize:
\begin{align}
    \mathcal{L}^{\mathrm{PIT}} = \frac{1}{TC} \sum_{t,c=1}^{T',C} \min_{\alpha \in \mathrm{Perm}(ct)} \mathcal{L}_{\alpha,tc}^\mathrm{ACCDOA}, \\
    \mathcal{L}_{\alpha, tc}^\mathrm{ACCDOA} = \frac{1}{N} \sum_{n=1}^{N} \mathrm{MSE} \left(\mathbf{P}_{\alpha,tcn}^*, \hat{\mathbf{P}}_{tcn}\right),
\end{align}
where $\alpha \in \mathrm{Perm}(t)$ is a possible frame-level permutation of $M$ tracks at the frame $t$
 and $\mathrm{Perm}(t)$ is a set of all possible permutations.
$\mathrm{MSE}(\cdot,\cdot)$ is the mean square error function.

\begin{table*}
    \centering
    \caption{Class-wise activitiy in the STARSS22 dataset~\cite{politis2022starss22}.}
    \vspace{-2mm}
    \label{tab:activity}
    \begin{tabular}{l|ccccccccccccc}
        \toprule
        & Fem. & Male & \multirow{2}{*}{Clap} & \multirow{2}{*}{Phone} & \multirow{2}{*}{Laugh} & Dom. & \multirow{2}{*}{Footsteps} &   \multirow{2}{*}{Door} & \multirow{2}{*}{Music} & Music. & \multirow{2}{*}{Faucet} & \multirow{2}{*}{Bell} & \multirow{2}{*}{Knock} \\
        & speech & speech & & & & sounds & & & & instr. & & & \\
        \midrule
        Frame coverage & \multirow{2}{*}{20.4} & \multirow{2}{*}{37.6} & \multirow{2}{*}{0.7} & \multirow{2}{*}{1.4} & \multirow{2}{*}{2.7} & \multirow{2}{*}{17.9} & \multirow{2}{*}{1.3} & \multirow{2}{*}{0.6} & \multirow{2}{*}{29.4} & \multirow{2}{*}{4.0} & \multirow{2}{*}{1.7} & \multirow{2}{*}{1.5} & \multirow{2}{*}{0.1} \\
        (\% total frames) & & & & & & & & & & & & & \\
        \bottomrule
    \end{tabular}
\end{table*}

\begin{table*}[t]
    \centering
    \caption{Evaluation result for the two datasets.}
    \vspace{-2mm}
    \label{tab:seldscore}
    \begin{tabular}{l|l|cccc|c}
        \toprule
        Fine-tuning dataset & Pretraining method & $ER_{\le 20^{\circ}} \downarrow$ & $F_{\le 20^{\circ}} \uparrow$ & $LE \downarrow$ & $LR \uparrow$ & $SELD \downarrow$
        \\
        \midrule
        \multirow{4}{*}{\vspace{-3mm}STARSS22$+$Synth}
        & None & 0.53 & 48.9\,\% & 18.2$^\circ$ & 68.7\,\% & 0.364 \\
        \cmidrule{2-7}
        & AVC\cite{arandjelovic2017look} & 0.52 & 49.7\,\% & 17.9$^\circ$ & 69.0\,\% & 0.359 \\
        \cmidrule{2-7}
        & AV-SSL with DOA-wise contrastive learning & \bf 0.51 & 50.5\,\% & 17.9$^\circ$ & \bf 70.1\,\% & 0.351 \\
        & AV-SSL with recording-wise contrastive learning & \bf 0.51 & \bf 51.6\,\% & \bf 17.1$^\circ$ & 69.0\,\% & \bf 0.349 \vspace{1mm} \\
        \toprule
        \multirow{4}{*}{\vspace{-3mm}STARSS22}
        & None & 0.65 & \bf 37.9\,\% & \bf 22.1$^\circ$ & \bf 58.4\,\% & \bf 0.452 \\
        \cmidrule{2-7}
        & AVC\cite{arandjelovic2017look} & \bf 0.63 & 37.7\,\% & 22.4$^\circ$ & 55.2\,\% & 0.458 \\
        \cmidrule{2-7}
        & AV-SSL with DOA-wise contrastive learning & 0.68 & 35.8\,\% & 22.9$^\circ$ & 53.7\,\% & 0.478 \\
        & AV-SSL with recording-wise contrastive learning & 0.67 & 36.3\,\% & 23.2$^\circ$ & 56.3\,\% & 0.467 \\
        \bottomrule
    \end{tabular}
\end{table*}

\begin{table*}[t]
    \centering
    \caption{Class-wise increase or decrease of the three SELD metrics by each pretraining.}
    \newcommand{\cP}[1] { \cellcolor{red!\xinttheiexpr (#1)*100\relax} \xintifboolexpr {#1<=0.3} {\textcolor{black}{$#1$}} {$#1$} }
    \newcommand{\cN}[1] { \cellcolor{blue!\xinttheiexpr (-#1)*100\relax} \xintifboolexpr {-#1<=0.3} {\textcolor{black}{$#1$}} {$#1$} }
    \vspace{-2mm}
    \label{tab:classws}
    \begin{tabular}{l|l|c|ccccccc}
        \toprule
        \multirow{2}{*}{Fine-tuning dataset} & \multirow{2}{*}{Pretraining method} & \multirow{2}{*}{$\Delta$ Metrics} & Dom. & \multirow{2}{*}{Door} & \multirow{2}{*}{Music} & Music. & \multirow{2}{*}{Bell} & \multirow{2}{*}{Knock} \\
        & & & sounds & & & instr. & & & \\
        \midrule
        \multirow{9}{*}{\vspace{-3mm}STARSS22$+$Synth}
        & \multirow{3}{*}{AVC~\cite{arandjelovic2017look}} 
        & $\Delta F_{\le 20^{\circ}}$ & \cP{+0.03} & \cN{-0.02} & \cN{-0.02} & \cP{+0.02} & \cP{+0.04} & \cP{+0.01} \\
        & & $\Delta (1 - LE/180)$ & \cP{+0.01} & \cP{+0.01} & \cN{-0.01} & {0.0} & \cP{+0.00} & \cN{-0.0} \\
        & & $\Delta LR$ & \cP{+0.02} & \cN{-0.01} & \cP{+0.05} & \cP{+0.01} & \cP{+0.08} & \cN{-0.0} \\
        \cmidrule{2-9}
        & \multirow{3}{*}{DOA-wise} 
        & $\Delta F_{\le 20^{\circ}}$ & \cP{+0.05} & \cP{+0.08} & \cP{+0.02} & \cN{-0.12} & \cP{+0.11} & \cN{-0.1} \\
        & & $\Delta (1 - LE/180)$ & \cP{+0.0} & \cP{+0.0} & \cN{-0.01} & \cN{-0.03} & \cP{+0.02} & \cN{-0.0} \\
        & & $\Delta LR$ & \cP{+0.07} & \cP{+0.07} & \cP{+0.09} & \cN{-0.02} & \cP{+0.01} & \cN{-0.05} \\
        \cmidrule{2-9}
        & \multirow{3}{*}{recording-wise} 
        & $\Delta F_{\le 20^{\circ}}$ & \cP{+0.1} & \cP{+0.06} & \cN{-0.04} & \cN{-0.01} & \cP{+0.12} & \cN{-0.03} \\
        & & $\Delta (1 - LE/180)$ & \cP{+0.01} & \cP{+0.01} & \cN{-0.02} & \cN{-0.0} & \cP{+0.02} & \cN{-0.0} \\
        & & $\Delta LR$ & \cP{+0.04} & \cN{-0.01} & \cP{+0.02} & \cN{-0.0} & \cP{+0.06} & \cN{-0.0} \\
        \toprule
        \multirow{9}{*}{\vspace{-3mm}STARSS22}
        & \multirow{3}{*}{AVC~\cite{arandjelovic2017look}} 
        & $\Delta F_{\le 20^{\circ}}$ & \cN{-0.02} & \cP{+0.09} & \cP{+0.0} & \cN{-0.05} & \cN{-0.26} & \cN{-0.18} \\
        & & $\Delta (1 - LE/180)$ & \cN{-0.01} & \cP{+0.01} & \cP{+0.0} & \cN{-0.01} & \cN{-0.09} & \cP{+0.01} \\
        & & $\Delta LR$ & \cP{+0.07} & \cP{+0.07} & \cN{-0.05} & \cP{+0.06} & \cN{-0.17} & \cN{-0.38} \\
        \cmidrule{2-9}
        & \multirow{3}{*}{DOA-wise} 
        & $\Delta F_{\le 20^{\circ}}$ & \cN{-0.0} & \cP{+0.04} & \cP{+0.02} & \cN{-0.14} & \cN{-0.2} & \cN{-0.07} \\
        & & $\Delta (1 - LE/180)$ & \cN{-0.0} & \cP{+0.0} & \cN{-0.01} & \cN{-0.05} & \cN{-0.06} & \cP{+0.02} \\
        & & $\Delta LR$ & \cP{+0.06} & \cP{+0.07} & \cN{-0.03} & \cP{+0.06} & \cN{-0.1} & \cN{-0.31} \\
        \cmidrule{2-9}
        & \multirow{3}{*}{recording-wise} 
        & $\Delta F_{\le 20^{\circ}}$ & \cP{+0.03} & \cP{+0.0} & \cN{-0.0} & \cN{-0.1} & \cN{-0.16} & \cN{-0.2} \\
        & & $\Delta (1 - LE/180)$ & \cN{-0.0} & \cP{+0.02} & \cP{+0.02} & \cN{-0.02} & \cN{-0.05} & \cN{-0.0} \\
        & & $\Delta LR$ & \cP{+0.05} & \cP{+0.14} & \cN{-0.14} & \cN{-0.06} & \cN{-0.09} & \cN{-0.46} \\
        \bottomrule
    \end{tabular}
\end{table*}

\section{Evaluation}

This section reports a comparative experiment
 conducted for evaluating the proposed AV-SSL methods
 with the two variants of contrastive learning.

\subsection{Network configuration}
\label{subsec:netconf}
The STFT spectrograms with 
 a shifting interval of $480$ samples
 and a window size of 
 1024  
 were converted to the mel spectrograms 
 with $F=64$ mel bins.
The number of DOAs $K$ was $220$.

\begin{figure}[t]
    \centering
    \includegraphics[width=.95\hsize]{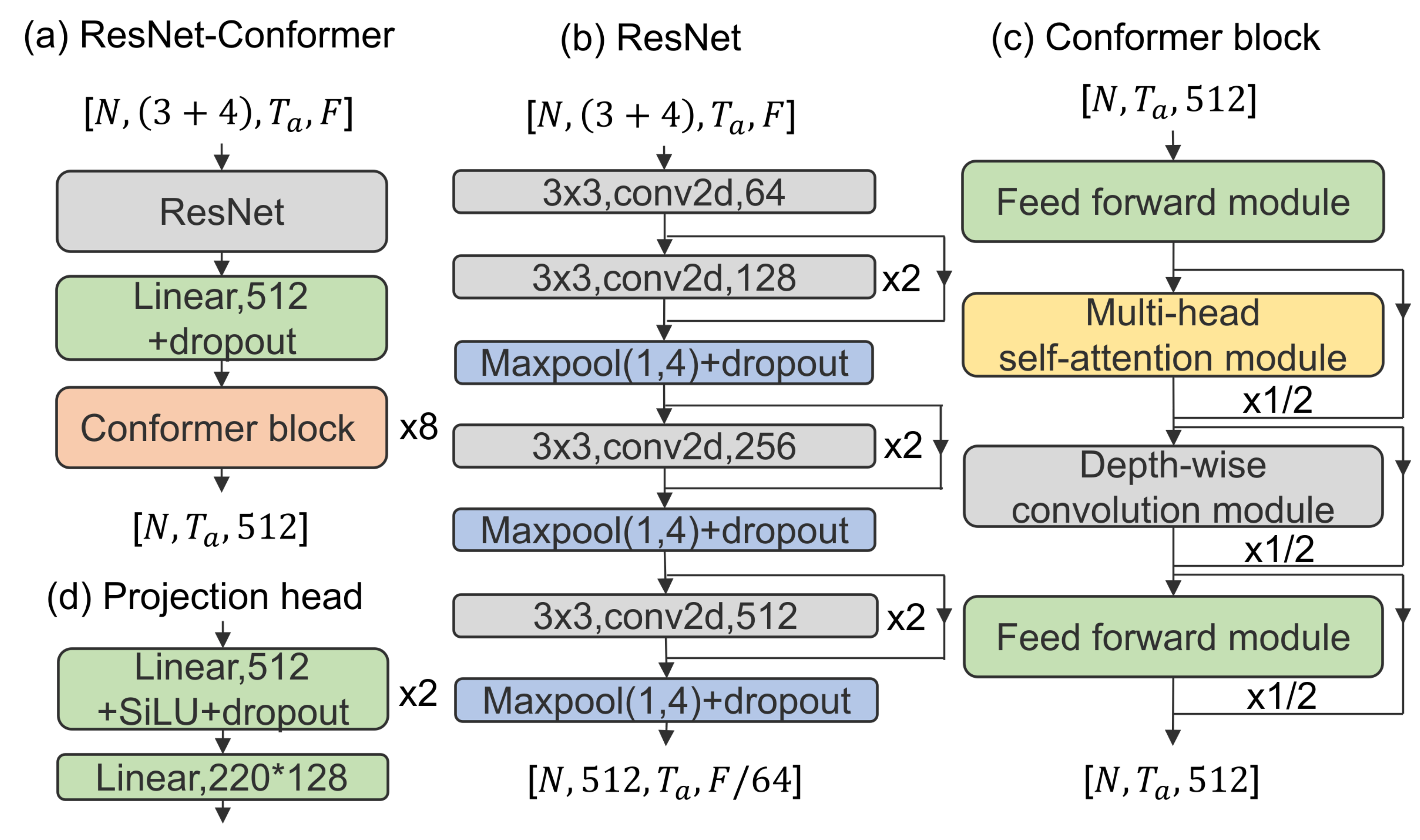}
    \caption{The architecture of the audio encoder.}
    \label{fig:resconf}
    \vspace{5mm}
    \includegraphics[width=.95\hsize]{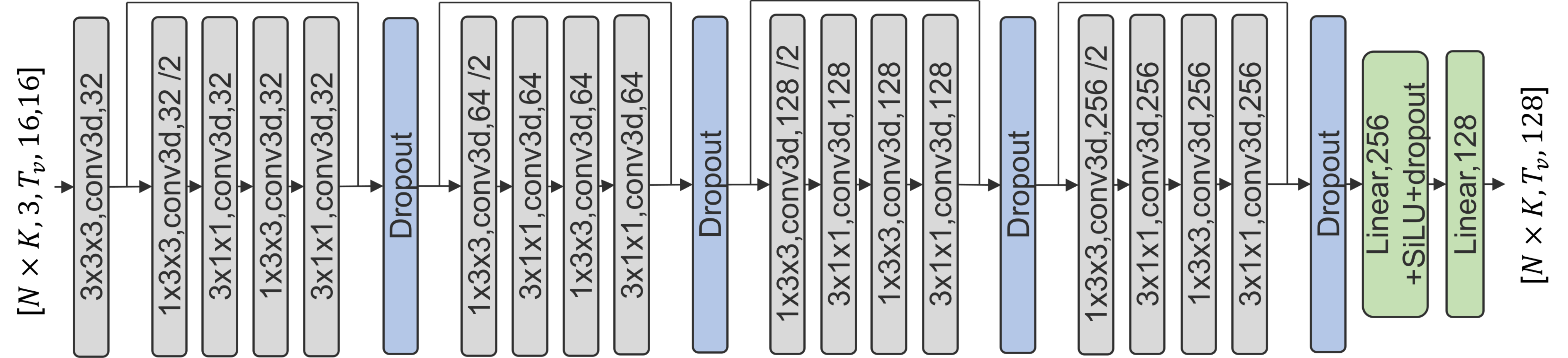}
    \caption{The architecture of the visual encoder.}
    \label{fig:r2p1d}
    \vspace{-3mm}
\end{figure}

We used a ResNet-Conformer~\cite{DBLP:journals/corr/abs-2101-02919} for the audio feature extractor $\mathcal{A}$ 
 as shown in Figs.~\ref{fig:resconf} (a)--(c).
The architecture of the projection head $\mathcal{H}$ consisted of 
 two linear+SiLU layers followed a linear layer
 which output dimension was $220 \times 128$
 as shown in Figs.~\ref{fig:resconf} (d).
The FOA audio data were resampled at a sampling rate of $24$ kHz 
 and split into $2$-second clips in the training.
  
The visual encoder $\mathcal{V}$
 consisted of $9$-layer $R(2+1)D$ convolution layers~\cite{perotin2018crnn} 
 as shown in Fig.~\ref{fig:r2p1d}.
The input equirectangular visual data were extracted 
 at the time resolution $T_v$ of $16$ ($8$ Hz).
The field of view $\psi$ of visual crops for each direction was $40^\circ$, 
 and the resolution $H \times W$ was $16 \times 16$.
The horizontal and vertical flipping was applied to the visual crops as data augmentation.  


\subsection{Pretraining}

We used the YT-360 dataset~\cite{morgado2020learning} 
 for pretraining.
It contains VR contents collected from YouTube, 
 each of which consists of a synchronized pair of FOA audio data 
 and equirectangular visual data, 
 including 246 hours of 10-second-long recordings 
 on diverse genres including music and sports.
After removing recordings with some missing channels, 
 104 hours of the training data 
 and 20 hours of the validation data 
 were used.


By the proposed AV-SSL methods,
 the ResNet-Conformer was trained for $100$ epochs
 using AdamW optimizer~\cite{loshchilov2017decoupled} 
 with a batch size of $4$, 
 a learning rate of $10^{-4}$, 
 a weight decay of $10^{-5}$.
Rotated recordings for the negative sample were obtained 
 by randomly rotating the equirectangular visual data 
 around the z-axis.
The temperature hyper-parameter $\tau$ was $0.1$.
The SpecAugment~\cite{park2019specaugment} 
 and the dropout with a rate of $0.1$ were used.
The model having the lowest validation loss was used for the downstream SELD task.

The conventional AV-SSL method
 called AVC \cite{arandjelovic2017look}
 was also evaluated as a baseline for comparison.
Specifically, the loss $\mathcal{L}_\mathrm{AVC}$ was 
 calculated with the recording-wise similarity $\mathcal{H}(\{\mathbf{z}^\audio_{it}\}_{t=1}^{T_v}, \{\mathbf{z}^\video_{it}\}_{t=1}^{T_v})$, 
 where $\mathbf{z}^\audio_{it} \in \mathbb{R}^{128}$ was 
 obtained by max-pooling the DOA-wise visual embeddings $\{ \mathbf{z}^\video_{ikt} \}_{k=1}^K$
 and $\mathbf{z}^\audio_{it} \in \mathbb{R}^{128}$ was 
 directly obtained from the output layer $\mathcal{H}$ 
 with output dimension $128$.

As in \cite{morgado2020learning}, 
 curriculum learning was introduced for the proposed methods, 
 where two audio feature extractors $\mathcal{A}$ were 
 first trained with AVC for $50$ epochs, 
 then trained with the two different proposed AV-SSLs for $50$ epochs.


\subsection{Fine-tuning}

The SELD performance was evaluated 
 on the STARSS22 and Synth1 datasets~\cite{politis2022starss22}. 
The STARSS22 is a dataset of actual recorded FOA data
 annotated with the activations, classes, 
 and DOAs of $13$ sound events 
 every $0.1$ seconds, 
 where the class labels follow 
 the Audioset ontology~\cite{gemmeke2017audio}.
This dataset consists of $2.9$ hours of training data 
 and 2 hours of validation data.
Note that some classes have very low frame coverages in STARSS22 
 as shown in Table~\ref{tab:activity}.
The Synth1 is a dataset of synthesized FOAs 
 annotated in the same way as STARSS22, 
 obtained by remixing source signals from FSD50K~\cite{fonseca2021fsd50k}
 with room RIRs from TAU-SRIR database~\cite{politis2022tau}.
Two datasets of different sizes constructed from these datasets
 were used for fine-tuning. 
The first one was a dataset made up of all training data 
 from STARSS22 and Synth1 (STARSS22+Synth1).
The second one consisted only of all training data from STARSS22 (STARSS22).
A dataset that consists of all validation data
 from STARSS22 and Synth1 was used as the validation dataset 
 in both conditions.
The original FOA data were split every 5 seconds,
 and they were sampled uniformly over all 13 classes
 to deal with the class imbalance in STARSS22.

Each pretrained model was fine-tuned 
 with a head $\mathcal{H}'$ 
 to output 3-track multi-ACCDOAs~\cite{shimada2022multi}.
The architecture of $\mathcal{H}'$ was 
 the same as the head $\mathcal{H}$
 shown in the Sec.~\ref{subsec:netconf}
 except that the output dimension was changed to $13 \times 3 \times 3$.
AdamW with a learning rate of $10^{-4}$ and a weight decay of $10^{-5}$ was used for training.
The models were trained for $1000$ epochs.
For data augmentation, the input FOA was randomly rotated.
The dropout rate was $0.1$ for ResNet-Conformer and $0.05$ for $\mathcal{H}'$.

Models pretrained with our methods were compared 
 with a non-pretrained model
 and a model pretrained with the conventional AVC method
 based on SELD score~\cite{politis2020overview} on the validation dataset.
The SELD score is obtained by averaging over four metrics as follows:
\begin{align}
    SELD = \frac{1}{4} \left( ER_{\le 20^{\circ}} +(1-F_{\le 20^{\circ}}) + \frac{LE}{180} + (1-LR) \right), \nonumber
\end{align}
 where the smaller score indicates better performance.
$ER_{\le 20^{\circ}}$ is the location-dependent error rate,
 where the prediction is counted as a true positive 
 only if the estimated class activity is correct 
 and the distance between the estimated and reference DOAs is smaller than $20^{\circ}$.
$F_{\le 20^{\circ}}$, $LE$ and $LR$ are the averages of class-wise metrics 
 $F_{c,\le 20^{\circ}}$, $LE_c$ and $LR_c$, respectively.
$F_{c,\le 20^{\circ}}$ is the location-dependent F1-score for class $c$. 
$LE_c$ is the average DOA estimation error over only the correctly detected events of class $c$.
$LR_c$ is the location-independent recall for class $c$.
$LE_c$ and $LR_c$ can be considered to be the performance of detection and localization, respectively.
The performance of each model was measured by averaging the metrics 
 over the epochs with the first to tenth highest SELD scores.

\subsection{Experimental results}

Table~\ref{tab:seldscore} shows 
 the overall SELD performances in terms of the five metrics, 
 i.e., $ER_{\le 20^{\circ}}$, $F_{\le 20^{\circ}}$, $LE$, $LR$, 
 and SELD score for each dataset and each pretraining method.
In the STARSS22+Synth1 dataset,  
 both the AV-SSL with DOA-wise contrastive learning 
 and the AV-SSL with recording-wise contrastive learning
 improved the SELD scores by about 1.5 pts,
 while the conventional AVC 
 improved only about 0.5 pts.
This result indicates that 
 our proposed approach is more suitable 
 for the pretraining of SELD than AVC.

Table~\ref{tab:classws} shows the performance gaps
 in the class-wise metrics, $F_{c,\le 20^{\circ}}$, $LE_c$ and $LR_c$, 
 between the non-pretrained model
 and each pretrained model,
 where all the metrics were transformed 
 to the range of $[0, 1]$ for readability,
 and only some of the classes required 
 for the following discussion
 were picked up due to the page limitation.
The proposed AV-SSLs degraded the SELD scores more than AVC
 when only a few hours of imbalanced data 
 were used for training (STARSS22).
This would be partly due to 
 the domain mismatch between YT-360 and STARSS22
 and the class imbalance of STARSS22.
As for the knock class, 
 which was not included in the pretraining dataset (YT-360),
 for example, 
 the detection performance $LR_c$ got particularly worse.
While in the classes related to home sounds 
 (e.g., domestic sounds and door),
 which frequently appear in YT-360,
 the detection performance was improved.

Another reason for the degradation would be
 that a large amount of background music data 
 independent from the visual data 
 are included in YT-360.
Background music was considered to have a negative impact 
 on the localization of sound events related to music
 because it did not spatially correspond to the paired visual data.
In fact, the localization performance $LE_c$ 
 was significantly degraded by the pretraining
 for the musical instruments and bell classes.



\section{Conclusion}

We proposed two variants of pretraining 
 an audio feature extractor useful for SELD 
 using spatial audio-visual recordings.
To obtain the latent audio features 
 representing not only the classes 
 but also the DOAs of sound events, 
 the audio encoder takes the FOA data as input, 
 and outputs the audio embeddings over the DOA grid.
The transfer learning with a sufficient amount of data
 showed the effectiveness of  
 the proposed AV-SSLs as pretraining for SELD.
For future work, 
 one should deal with the deterioration in the SELD performance
 when sufficient labeled data is unavailable.
One of the promising approaches would be to prepare the dataset of spatial audio-visual recordings 
 covering various domains with good correspondence between audio and visual data.
 
 .

\section*{Acknowledgment}
This work was supported in part by NEDO, JSPS KAKENHI Nos.~20H00602, 19H04137, and ISHIZUE 2023 of Kyoto University.

\printbibliography

\end{document}